\documentclass[12pt]{article}
\usepackage{graphicx}
\usepackage{epstopdf}
\usepackage{amsfonts}
\usepackage{amssymb}
\usepackage[footnotesize]{caption2}
\usepackage{mathrsfs}
\usepackage{amsmath}
\usepackage[numbers, sort&compress]{natbib}

\usepackage{authblk}

\begin{document}
\renewcommand{\thefootnote}{\fnsymbol{footnote}}
\title{Remnants of black holes from rainbow gravity in terms of a new VSL theory}

\author{\small Ying-Jie Zhao\thanks{E-mail address: xiangyabaozhang@qq.com} , 
\small Xiaojie Li\thanks{E-mail address: 13256199291@163.com} ,
\small Guang-Rui Yao\thanks{E-mail address: coldstar\_ygr@qq.com} ,
\small Yan-Fang Ji\thanks{E-mail address: jiyanfang88@qq.com} ,
\small Nisar Ahmad\thanks{E-mail address: nisar@mail.ustc.edu.cn}}

\affil{\small Qilu Institute of Technology, 3028 Jingshi East Road, Jinan City, Shandong Province, China}

\renewcommand{\Authands}{, }
\date{}
\maketitle
\begin{abstract}

\setlength{\parindent}{0pt} \setlength{\parskip}{1.5ex plus 0.5ex minus 0.2ex} 
\noindent
The gravity's rainbow function is derived in terms of a new varying speed of light (VSL) theory that varying velocity of light in the rainbow gravity becomes smaller when the energy of photons increases. In light of the new theory we calculate the modified temperature, entropy and heat capacity of the Schwarzschild, Kerr, AdS black holes, as well as spinning black rings. Our results demonstrate that in rainbow gravity the behaviors of various black holes have remarkably essential difference from those of standard black holes near the Planck scale, and a logarithmic term emerges in the formula of the entropy of every black object. 

\vskip 10pt
\noindent
{\bf PACS Number(s)}: 04.60.-m, 04.70.Dy,  04.50.Kd
\vskip 5pt
\noindent

\end{abstract}

\thispagestyle{empty}

\newpage

\setcounter{page}{1}

\section{Introduction}

Gravity's rainbow is a remarkable attempt to reach the unification of general relativity and quantum mechanics, which is a goal of high difficulty due to ultraviolet-divergence and non-renomalizability of the existing quantum gravity theories. Various candidate theories of quantum gravity have been established such as string theory \cite{STRINGY,FATE,SUPERSTRING_1,SUPERSTRING_2,SUPERSTRING_3,MINIMUM}, loop quantum gravity \cite{LOOP}, quantum geometry \cite{QUANTUM}, spacetime discreteness \cite{DISCRETENESS}, spacetime foam models \cite{FOAM}, \textit{etc.}. They demonstrate a common feature that Lorentz violation is indispensable when building up an appropriate theory of quantum gravity; that is, it is essential to modify the standard energy-momentum dispersion relation into a new form.

A promising approach to achieve this aim is to generalize the special relativity to a double special relativity (DSR) \cite{DSR} that contains two fundamental constants, the velocity of light and the Planck energy. The simplest DSR based on a nonlinear Lorentz transformation in momentum space reveals that the usual dispersion relations of special relativity must be corrected at Planck scale. The theory of DSR could be extended to a curved manifold to deliver information of gravity. In such a framework the spacetime metric is no longer a freely chosen one but depends on the energy of a probing particle; the spacetime geometry in consequence leads to a parameter family of energy dependent metrics, known as the gravity's rainbow \cite{GRAVITY}.

The modified dispersion relations in  gravity's rainbow can be collectively expressed in a form: \quad ($\hbar=c=k_B=1$ hereinafter)
\begin{eqnarray}
E^2 f^2\left(\frac{E}{E_p}\right)-p^2  g^2\left(\frac{E}{E_p}\right) = m^2,
\end{eqnarray}
where $E$ represents the energy of a probing particle, and $E_p$ the Planck energy. The two functions $f\left(E/{E_p}\right)$ and $g\left(E/{E_p}\right)$ have the following behaviors in the infrared limit $E/E_p \rightarrow 0$:
\begin{eqnarray}
\lim_{E/{E_p} \to 0}{f\left(\frac{E}{E_p}\right)}= 1,  \quad \lim_{E/{E_p} \to 0}{g\left(\frac{E}{E_p}\right)}= 1.
\end{eqnarray}
According to Magueijo and Smolin \cite{GRAVITY}, the idea of energy-dependent spacetime metric for gravity's rainbow is realized via the following modification:
\begin{eqnarray}
g\left( E \right) = \eta^{ab} e_a\left( E \right) \otimes e_b\left( E \right).
\end{eqnarray}
The frame fields depending on energy $E$ are given by
\begin{eqnarray}
{e_0} \left( E \right) =  \frac{\tilde{e}_0}{f\left(E/{E_p}\right)}, \quad  {e_i} \left( E \right) =  \frac{\tilde{e}_i}{g\left(E/{E_p}\right)},
\end{eqnarray}
where $\sim$ refers to the energy-independent frame fields. The metric in flat spacetime reads
\begin{eqnarray}
ds^2=  -\frac{dt^2}{f\left(E/{E_p}\right)^2}+\frac{d{x_i}d{x^i}}{g\left(E/{E_p}\right)^2}.
\end{eqnarray}
The modified temperature of the black hole in gravity's rainbow is calculated as
\begin{eqnarray}
T=  \frac{g\left(E/{E_p}\right)}{f\left(E/{E_p}\right)} {T_0},
\end{eqnarray}
where $T_0$ is the standard temperature of black hole. The Einstein's equations are therefore modified as
\begin{eqnarray}
G_{\mu\nu}\left(\frac{E}{E_p}\right) =8\pi G T_{\mu\nu}\left(\frac{E}{E_p}\right),
\end{eqnarray}
$G_{\mu\nu}$ and $T_{\mu\nu}$ being functions of $E/{E_p}$.

The rainbow functions $f\left(E/{E_p}\right)$ and $g\left(E/{E_p}\right)$ have various types, giving rise to different modifications of the dispersion relation. For instance, the most
studied rainbow function proposed by Amelino-Camelia reads \cite{DISTANCE, LIVING}:
\begin{eqnarray}
f\left(\frac{E}{E_p}\right) = 1,  \quad g\left(\frac{E}{E_p}\right) = \sqrt{1 - \eta \left(\frac{E}{E_p}\right)^n} ,
\end{eqnarray}
which can be obtained from loop quantum gravity and noncommutative versions of Minkowski spacetime \cite{LIVING}. The second example is a function compatible with the results from the hard spectra of gamma-ray bursts at cosmological distance \cite{FOAM,DSR,PLANCKIAN}:
\begin{eqnarray}
f\left(\frac{E}{E_p}\right) = \frac{e^{\beta E/{E_p}}-1}{\beta E/{E_p}},  \quad g\left(\frac{E}{E_p}\right) = 1.
\end{eqnarray}
The third example is a function in which the velocity of the light keeps as a constant \cite{LORENTZ}
\begin{eqnarray}
f\left(\frac{E}{E_p}\right) = \frac{1}{1-\lambda E/{E_p}},  \quad g\left(\frac{E}{E_p}\right) = \frac{1}{1-\lambda E/{E_p}}.
\end{eqnarray}

Recent applications of rainbow gravity to investigation of  black holes have attracted much attention and several achievements have been made \cite{ALI1,ALI2,HENDI1,HENDI2,KUMAR,ANACLETO,SANTOS,BAKKE,MONTIGNY,TUDESHKI,PANAH}. For instance, a result of standard black hole thermodynamics is that a small black hole of the Planck scale radiates continuously such that the black hole temperature rises to infinity until the black hole mass decreases to zero \cite{HAWKING_1,HAWKING_2}. For this paradox, the gravity's rainbow provides a natural restriction that the horizon radius of a black hole cannot be less than the scale of the Planck length by the end of the evaporation process, while the remnant of the black hole at the final stage of evaporation has zero entropy, zero heat capacity and finite temperature. The entropy of a black hole does not obey the area theorem in a strict manner, but contains an additional correction. Various studies focusing on the effects of gravity's rainbow functions on the thermodynamics of black holes, seen in Refs.\cite{KIM,DEHGHANI,MORAIS,HAMIL}, have been made. Moreover, the energy-dependent functions of  gravity's rainbow have influenced several properties in other areas, such as modified gravity, astrophysics and cosmology. 

Based on the above gravity's rainbow models, we propose in this paper a new form of VSL theory (Varying Speed of Light):
\begin{eqnarray}
f\left(E/{E_p}\right) = \frac{1}{1-\eta E/{E_p}},  \quad g\left(E/{E_p}\right) = 1,
\end{eqnarray}
where $\eta$ is a dimensionless positive constant. This leads to the following new energy-momentum relation
\begin{eqnarray}
\frac{E^2}{\left( 1-\eta  \frac{E}{E_p}\right)^2} - p^2 = m^2 ,
\end{eqnarray}
where the maximal energy reaches $\frac{E_p}{\eta}$. For massless particles, $m=0$, the dispersion relation turns to be
\begin{eqnarray}
p = \frac{E}{1-\eta \frac{E}{E_p}},
\end{eqnarray}
hence a spacetime with the energy-dependent velocity is achieved,
\begin{eqnarray}
\tilde{c} = \left(1-\eta\frac{E}{E_p}\right)^2.
\end{eqnarray}
In this model, varying velocity of light in the rainbow gravity becomes smaller when the energy of photons increases.

In the coming Sections 2--5, emphasis will be placed on the above rainbow gravity, especially its impact on black hole and black ring thermodynamics. We will show that differences between the situations \textit{with} and \textit{without} rainbow gravity effects, for Schwarzschild black hole, Kerr black hole, AdS black hole and spinning black ring, respectively. Finally a brief conclusion will be presented in Section 6.
\\
\\

%%%%%%%%%%%%%%%%%%%%%%%%%%%%%%%%%%%%%%%%%%%%%%%%%%%%%%%%%%
%%%%%%%%%%%%%%%%%%%%%%%%%%%%%%%%%%%%%%%%%%%%%%%%%%%%%%%%%%
%%%%%%%%%%%%%%%%%%%%%%%%%%%%%%%%%%%%%%%%%%%%%%%%%%%%%%%%%%

\section{Schwarzschild black hole}

The modified metric of gravity's rainbow reads \cite{GRAVITY}
\begin{eqnarray}
ds^2 = -\frac{1-\frac{2M}{r}}{f^2(E/{E_p})}dt^2+\frac{dr^2}{\left(1-\frac{2M}{r}\right)g^2(E/{E_p})}+\frac{r^2 d\Omega}{g^2(E/{E_p})}.
\end{eqnarray}
It is assumed that in a Hawking radiation process a particle is emit from a Schwarzschild black hole.
According to the near-horizon geometry, the measurement precision of the particle's position is of the order of the event horizon radius, $\Delta x \approx r_+$.  In the light of the Heisenberg's uncertainty principle,
\begin{eqnarray}
\Delta p \geq 1/{\Delta x},
\end{eqnarray}
we deduce the relation between the event horizon radius of the Schwarzschild black hole and the momentum of the particle emission in the Hawking radiation,
\begin{eqnarray}
\Delta p \geq 1/{\Delta x} \approx 1/{r_+}.
\end{eqnarray}
This relation can be further translated into a bound put upon the energy \cite{ADLER, CAVAGLIA, MEDVED, ARZANO}
\begin{eqnarray}
E \geq 1/{\Delta x} \approx 1/{r_+},
\end{eqnarray}
where $E$ is understood as the particle energy suppressed by the Planck energy $E_p$.
Then, linking the surface gravity directly to the black hole temperature \cite{ALI1}, we have
\begin{eqnarray}
T=\frac{\kappa}{2\pi}=\frac{g(E/{E_p})}{f(E/{E_p})}\frac{1}{8\pi M},  \label{tem1}
\end{eqnarray}
and therefore achieve the corrected temperature in terms of the  event horizon radius,
\begin{eqnarray}
T=\frac{1}{4\pi r_+}(1-\frac{\eta E }{ E_p}) = \frac{1}{4\pi r_+}(1-\frac{\eta }{r_+ E_p}),
\end{eqnarray}
where $r_+=2M$.

It is seen that the temperature goes to zero when the black hole evaporates to a small one with horizon radius
\begin{eqnarray}
r_+ = \frac{\eta}{E_p}.
\end{eqnarray}
This corresponds to a minimal mass
\begin{eqnarray}
M_{min} =  \frac{\eta}{2 E_p}.
\end{eqnarray}
Figures 1, 2 and 3 demonstrate the modified temperatures, entropies and heat capacities, respectively, in the units $\eta =1, E_p=1$. Figure 1 shows that, in the framework of rainbow gravity, the scale of the black hole cannot be less than the Planck mass by the end of the evaporation process, while the black hole's remnant at the final stage of evaporation has a zero temperature.

\begin{figure}[!htbp]
\centering
\includegraphics[height=6cm]{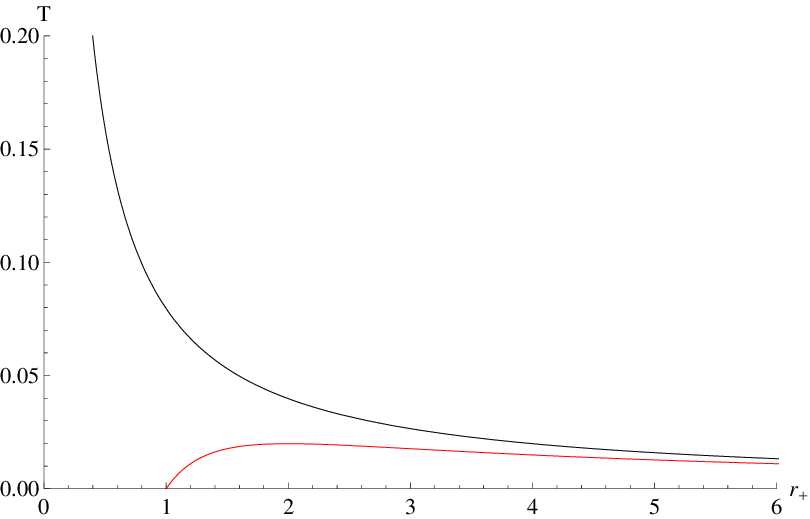}
\caption{{\footnotesize The standard (black) and modified temperature (red) versus the horizon radius for a Schwarzschild black hole.}}
\end{figure}

The entropy can be derived from the first law of black hole thermodynamics
\begin{eqnarray}
S= \int{\frac{dM}{T}} = \pi{r_+}^2 + \frac{2\pi \eta}{{E_p}} {r_+} +\frac{2 \pi \eta^2}{{E_p}^2} \ln{\left({r_+} - \frac{\eta}{E_p} \right)},
\end{eqnarray}
which degenerates to $S=\pi{r_+}^2$ when $\eta\rightarrow 0$. We plotted this result in Figure 2. The reason why negative values of the entropy emerge near the minimal horizon radius (seen also in the following sections) lies in that the real minimal mass of the black hole is not exactly equal to the $M_{min}$ we have deduced, but a little bit larger than that; or, at an ultrashort distance there exists new physics unknown to us so far. 

\begin{figure}[!htbp]
\centering
\includegraphics[height=6cm]{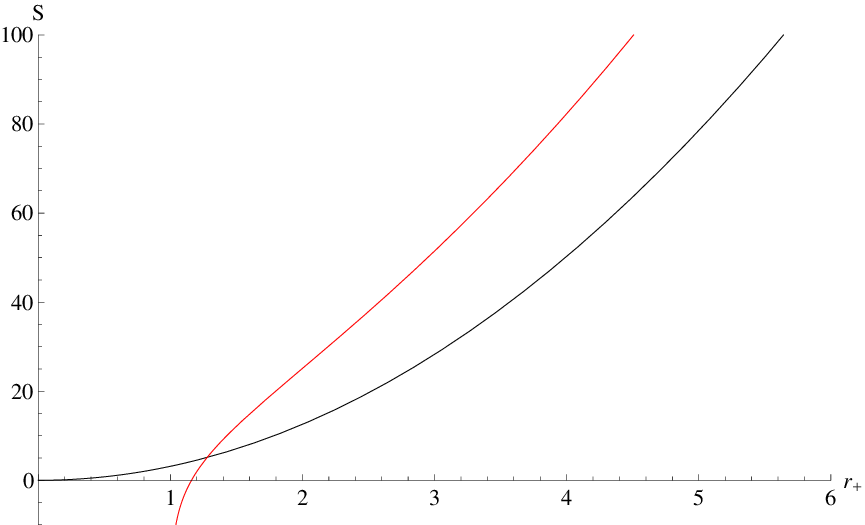}
\caption{{\footnotesize  The standard (black) and modified entropy (red) versus the horizon radius for a Schwarzschild black hole.}}
\end{figure}

The heat capacity of the black hole is determined by
\begin{eqnarray}
C =\frac{\partial M}{\partial T}= \frac{2 \pi E_p {r_+}^3}{2\eta - r_+ E_p},
\end{eqnarray}
showing that the heat capacity $C_J>0$ for a small horizon radius, and $C_J<0$ for a large one. Figure 3 indicates that the heat capacity diverges as the temperature reaches its maximum value, and decreases to zero when the horizon radius reaches its minimal value.
\begin{figure}[!htbp]
\centering
\includegraphics[height=6cm]{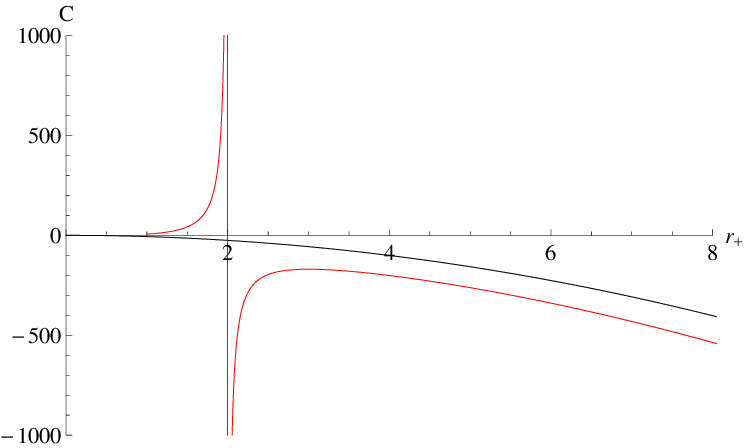}
\caption{{\footnotesize The standard (black) and modified heat capacity (red) versus the horizon radius for a Schwarzschild black hole.}}
\end{figure}

%%%%%%%%%%%%%%%%%%%%%%%%%%%%%%%%%%%%%%%%%%%%%%%%%%%%%%%%%%
%%%%%%%%%%%%%%%%%%%%%%%%%%%%%%%%%%%%%%%%%%%%%%%%%%%%%%%%%%
%%%%%%%%%%%%%%%%%%%%%%%%%%%%%%%%%%%%%%%%%%%%%%%%%%%%%%%%%%

\section{Kerr black hole}

The metric of Kerr black hole is given by
\begin{eqnarray}
ds^2 &=& - dt^2+ \frac{2Mr}{\Sigma }\left(dt - {a \sin^2\theta d\phi}\right)^2 + \frac{\Sigma dr^2}{\Delta}    \nonumber  \\
&&+ {\Sigma d\theta^2+(r^2 +a^2)\sin \theta^2 d\phi^2},
\end{eqnarray}
where
\begin{eqnarray}
\Sigma = r^2 + a^2 \cos^2 \theta, \quad  \quad \Delta =  r^2 + a^2 - 2Mr.
\end{eqnarray}
The temperature of Kerr black hole can be found in Ref.\cite{MA},
\begin{eqnarray}
T_0 = \frac{{r_+}^2 - a^2}{4 \pi {r_+}({r_+}^2 + a^2)}
\end{eqnarray}

As Ahmed Farag Ali {\em et al.} revealed in Ref.\cite{KERR}, in gravity's rainbow we simply make the changes $dt \rightarrow dt/f\left(E/{E_p}\right)$  and $d{x^i} \rightarrow d{x^i}/g\left(E/{E_p}\right)$ to get the modified metric of Kerr black hole. 
Using the above method we achieve the corrected temperature in gravity's rainbow, which is expressed by the horizon radius of the black hole $r_+$ as well as the angular momentum per unit mass, $a = J/M$. Here we employ $E  \approx 1/{r_+} \approx \sqrt{4\pi/A}$, where $A$ denotes the area of the outer event horizon. It takes $A=4\pi {r_+}^2$ for the Schwarzschild case, but $A = 4\pi ({r_+}^2 + a^2)$ for the Kerr case,
\begin{eqnarray}
T&&=\frac{g(E/{E_p})}{f(E/{E_p})} T_0 \nonumber \\
&&=\frac{{r_+}^2 - a^2}{4\pi {r_+}({r_+}^2+a^2)}\left(1-\frac{\eta}{E_p} \sqrt{\frac{4\pi}{A}}\right) \nonumber \\
&&= \frac{{r_+}^2 - a^2}{4\pi {r_+}({r_+}^2+a^2)}\left(1-\frac{\eta}{E_p \sqrt{{r_+}^2+a^2}}\right),
\end{eqnarray}
\begin{figure}[!htbp]
\centering
\includegraphics[height=6cm]{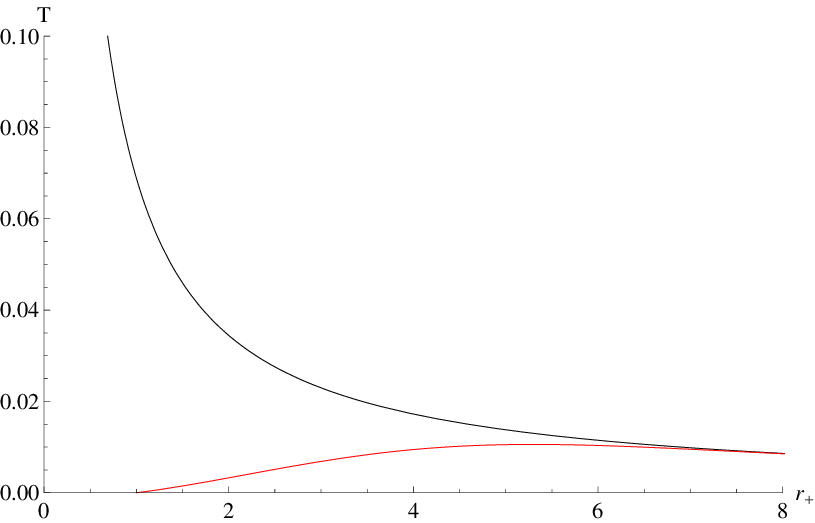}
\caption{{\footnotesize The standard (black) and modified temperature (red) versus the horizon radius for a Kerr black hole.}}
\end{figure}
with the horizon radius deduced from $M =\frac{{r_+}^2+a^2}{2 r_+}$.

The first law of the black hole thermodynamics leads to the entropy
\begin{eqnarray}
S &&= \int\left(\frac{dM}{T}-\frac{\Omega dJ}{T}\right)     \nonumber     \\
&&= \pi \left( {r_+}^2 + a^2 \right) + \frac{2\pi \eta}{E_p} {\sqrt{{r_+}^2+a^2}}+\frac{2\pi \eta^2}{{E_p}^2} \ln\left({\sqrt{{r_+}^2+a^2}-\frac{\eta}{E_p}}\right),
\end{eqnarray}
\begin{figure}[!htbp]
\centering
\includegraphics[height=6cm]{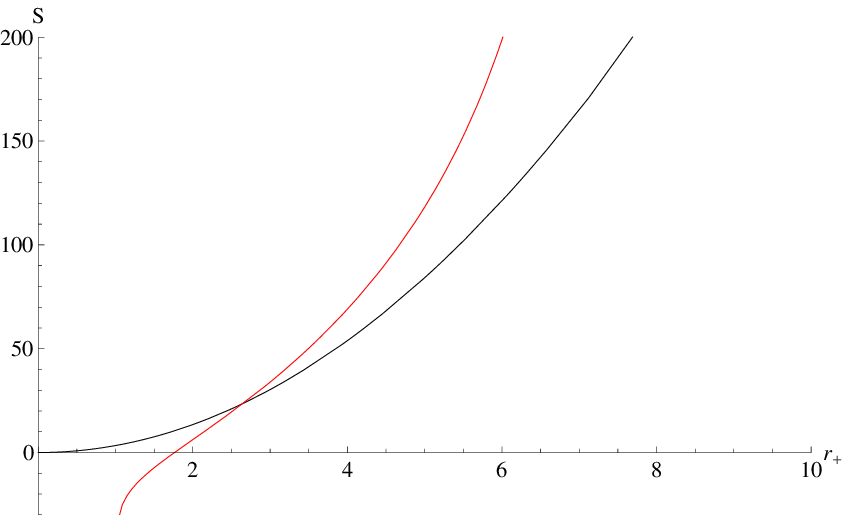}
\caption{{\footnotesize The standard (black) and modified entropy (red) versus the horizon radius for a Kerr black hole.}}
\end{figure}
where the angular velocity $\Omega$ and the angular momentum $J$ are given by, respectively,
\begin{eqnarray}
\Omega = \frac{a}{{r_+}^2+a^2}, \text{ \ \ \ \ \ \ \ }  J = \frac{a(a^2+{r_+}^2)}{2 r_+}.
\end{eqnarray}
This entropy degenerates to $S=\pi({r_+}^2+a^2)$ when $\eta\rightarrow 0$.
Similarly, the heat capacity determining the stability of black holes at a constant angular momentum is calculated as \cite{PERRY}
\begin{eqnarray}
C_J &=& T \left( \frac{\partial S}{\partial T}\right)_J   \nonumber     \\
& =& {\frac{\partial (S,J)}{\partial(r_+, a)}}{\bigg /}{\frac{\partial (T,J)}{\partial(r_+, a)}}    \nonumber            \\
& =&\scalebox{1.1} {$ \frac{2\pi ({r_+}^2-a^2)({r_+}^2+a^2)^{\frac{5}{2}}}{a^4 (3\sqrt{{r_+}^2+a^2}-4\eta {E_p}^{-1})-{r_+}^4(\sqrt{{r_+}^2+a^2}-2\eta {E_p}^{-1})+6 a^2 {r_+}^2 (\sqrt{{r_+}^2+a^2}-\eta{E_p}^{-1})}$}.
\end{eqnarray}
\begin{figure}[!htbp]
\centering
\includegraphics[height=6cm]{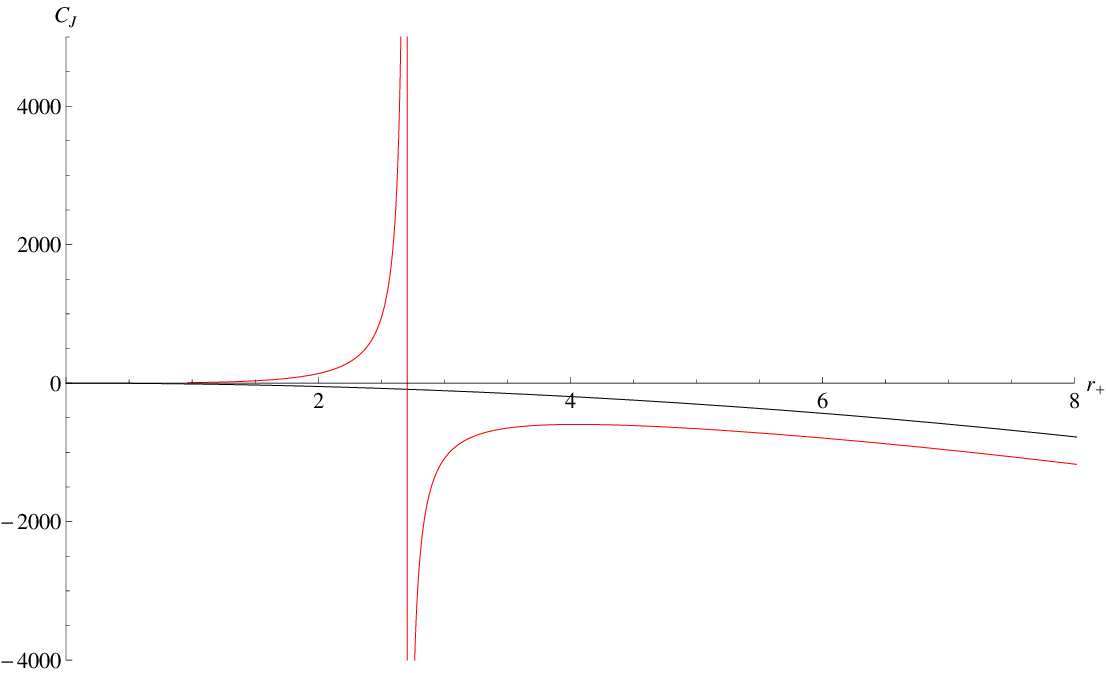}
\caption{{\footnotesize The standard (black) and modified heat capaity (red) versus the horizon radius for a Kerr black hole.}}
\end{figure}

Figure 4, 5 and 6 in the units $\eta = 1, E_p = 1$ and $a = M/2$ indicate that the modified temperature, entropy, and heat capacity of a Kerr black hole are similar to those of a Schwarzschild black hole. Besides two reasons mentioned in Section 2, another reason why the entropy of Kerr black hole declines to negative values near $r_+ \approx 2$ should be the approximation $E \approx 1/r_+ \approx \sqrt{4\pi/A}$ we made above.

%%%%%%%%%%%%%%%%%%%%%%%%%%%%%%%%%%%%%%%%%%%%%%%%%%%%%%%%%%
%%%%%%%%%%%%%%%%%%%%%%%%%%%%%%%%%%%%%%%%%%%%%%%%%%%%%%%%%%
%%%%%%%%%%%%%%%%%%%%%%%%%%%%%%%%%%%%%%%%%%%%%%%%%%%%%%%%%%

\section{AdS black hole}

The metric of AdS black holes in rainbow gravity is given by Ref.\cite{YEKTA}
\begin{eqnarray}
ds^2 = -\frac{F(r)}{f^2(E/{E_p})}dt^2 + \frac{dr^2}{F(r) g^2(E/{E_p})} +\frac{r^2 d\Omega^2}{g^2(E/{E_p})},
\end{eqnarray}
with
\begin{eqnarray}
F(r)= 1- \frac{2M}{r}+\frac{r^2}{l^2},
\end{eqnarray}
where $l$ is the radius of a charged AdS black hole. The temperature can be obtained by using $E \approx 1/{r_+} $
\begin{eqnarray}
T=\frac{3{r_+}^2+l^2}{4 \pi l^2 {r_+}}\left(1-\frac{\eta}{r_+ E_p}\right).
\end{eqnarray}
\begin{figure}[!htbp]
\centering
\includegraphics[height=6cm]{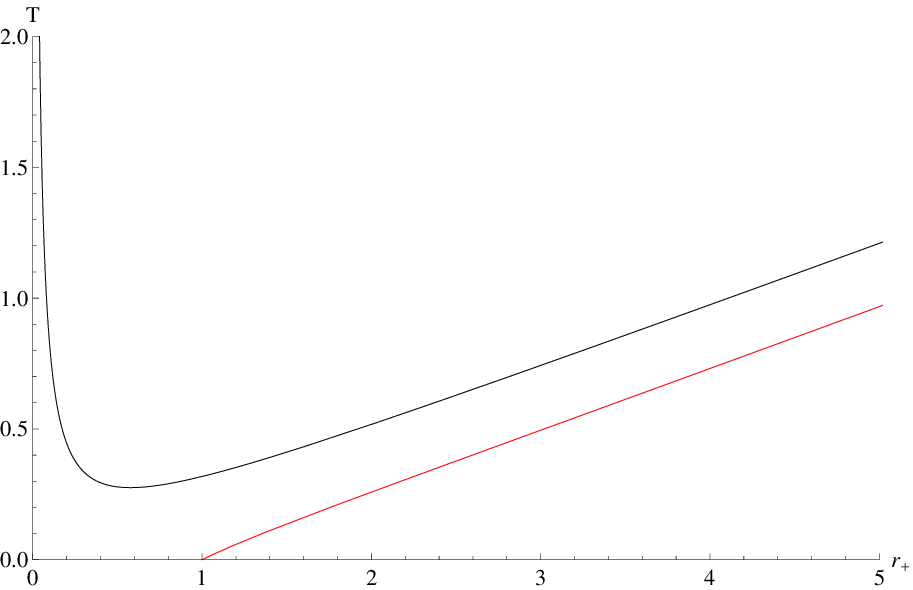}
\caption{{\footnotesize The standard (black) and modified temperature (red) versus the horizon radius for an AdS black hole.}}
\end{figure}
Figure 7, 8 and 9 are the plots of the modified temperatures, entropies and heat capacities in the units $\eta =1, E_p=1$ and $l=1$.
From Figure 7 we can see that the standard temperature of an AdS black hole has a positive minimal value $\frac{2\sqrt{3}}{4\pi l}$, while its modified temperature reduces to zero at the point of minimal horizon radius $\eta/{E_p}$.
Moreover, according to $F(r_+) = 0$, the mass turns to be $M = \frac{{r_+}^3 + l^2 {r_+}}{2 l^2}$, which leads to the minimal mass
\begin{eqnarray}
M_{min} = \frac{\eta \left(\eta^2 +l^2 {E_p}^2 \right)}{2 l^2 {E_p}^3}.
\end{eqnarray}
The entropy is derived in the same way,
\begin{eqnarray}
S=\int\frac{dM}{T}= \pi {r_+}^2+\frac{2\pi \eta}{E_p} {r_+}+\frac{2\pi \eta^2}{{E_p}^2}\ln \left(r_+ - \frac{\eta}{E_p} \right),
\end{eqnarray}
 \begin{figure}[!htbp]
\centering
\includegraphics[height=6cm]{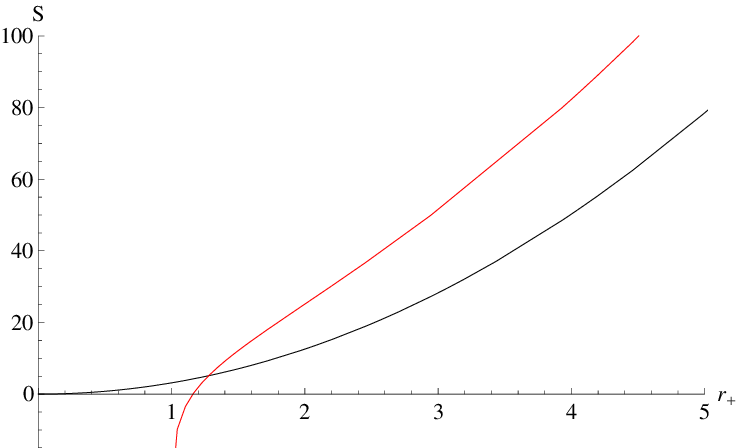}
\caption{{\footnotesize The standard (black) and modified entropy (red) versus the horizon radius for an AdS black hole.}}
\end{figure}
and the heat capacity of an AdS black holes is obtained as
\begin{eqnarray}
C = \frac{\partial M}{\partial T} = \frac{2\pi {r_+}\left(3{r_+}^4+l^2{r_+}^2\right)}{3{r_+}^3-l^2\left(r_+ - 2 \eta {E_p}^{-1}\right)}.
\end{eqnarray}
Figure 9 indicates that, in contrast with the above two black holes, the modified heat capacity here has a positive minimal value $2\pi (\eta/{E_p})^2$ at the point of minimal horizon radius $r_+=\frac{\eta}{E_p}$; meanwhile the standard temperature
diverges at a point  $r_+ = \frac{l}{\sqrt{3}}$ and decreases to zero at $r_+=0$.
\begin{figure}[!htbp]
\centering
\includegraphics[height=6cm]{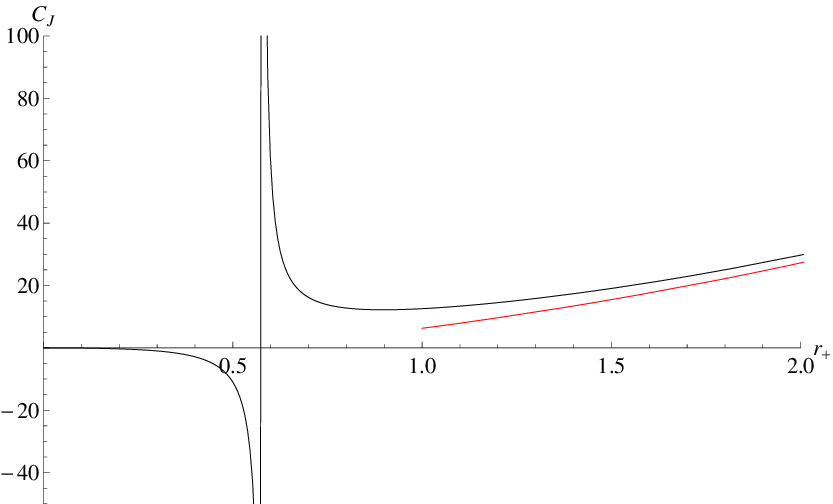}
\caption{{\footnotesize  The standard (black) and modified heat capaicty (red) versus the horizon radius for an AdS black hole.}}
\end{figure}

%%%%%%%%%%%%%%%%%%%%%%%%%%%%%%%%%%%%%%%%%%%%%%%%%%%%%%%%%%
%%%%%%%%%%%%%%%%%%%%%%%%%%%%%%%%%%%%%%%%%%%%%%%%%%%%%%%%%%
%%%%%%%%%%%%%%%%%%%%%%%%%%%%%%%%%%%%%%%%%%%%%%%%%%%%%%%%%%

\section{Spinning black ring}

The metric of a spinning black ring in five dimensions takes the form \cite{ALI2, PERRY, EMPARAN, ZHAO}
\begin{eqnarray}
ds^2 &=&  -\frac{A dt^2}{f^2(E/{E_p})}+\frac{dy^2}{Bg^2(E/{E_p})}+g_{\psi\psi}\left[\frac{d\psi}{g(E/{E_p})}+\frac{N^\psi dt}{f(E/{E_p})}\right]^2+\frac{g_{xx} dx^2}{g^2(E/{E_p})}     \nonumber \\
& & +\frac{g_{\phi\phi} d\phi^2}{g^2(E/{E_p})},
\end{eqnarray}
with
\begin{eqnarray}
A(x,y) &\equiv& \frac{F(y)}{F(x)}\left[1-\frac{C(\nu,\eta)^2 (1+y)^2}{F(x)^2 G(y)/(x-y)^2+C(\nu,\eta)^2(1+y)^2} \right], \\
B(x,y) &\equiv& -\left[\frac{R^2 F(x)}{(x-y)^2 G(y)}\right]^{-1},
\end{eqnarray}
\begin{eqnarray}
F(x)\equiv 1+\eta x, \quad G(x)\equiv(1-x^2)(1+\nu x),  \quad
C(\nu,\lambda)\equiv \sqrt{\eta(\eta-\nu)\frac{1+\eta}{1-\eta}}.
\end{eqnarray}
The dimensionless parameter $\nu$ represents the shape of the horizon, and $R$ a scale factor.
Hence we can deduce the modified temperature of spinning black ring with the relation
 $E \geq 1/{\Delta x \approx 1/({\nu R})} $ \cite{ALI2}
\begin{eqnarray}
T = \frac{(1-\nu)\sqrt{1+\nu^2}}{4\sqrt{2}\pi\nu R }\left(1-\frac{\eta}{\nu R E_p}\right).
\end{eqnarray}
We plot the curves of the modified temperatures, entropies and heat capacities versus the dimensionless parameter $\nu$ in Figure 10, 11 and 12, in the units $\eta=1, E_p = 1$ and $R =5$. Thus Figure 10 reveals a fact that the modified temperature goes to zero at $\nu = \frac{\eta}{R E_p}$.
\begin{figure}[!htbp]
\centering
\includegraphics[height=6cm]{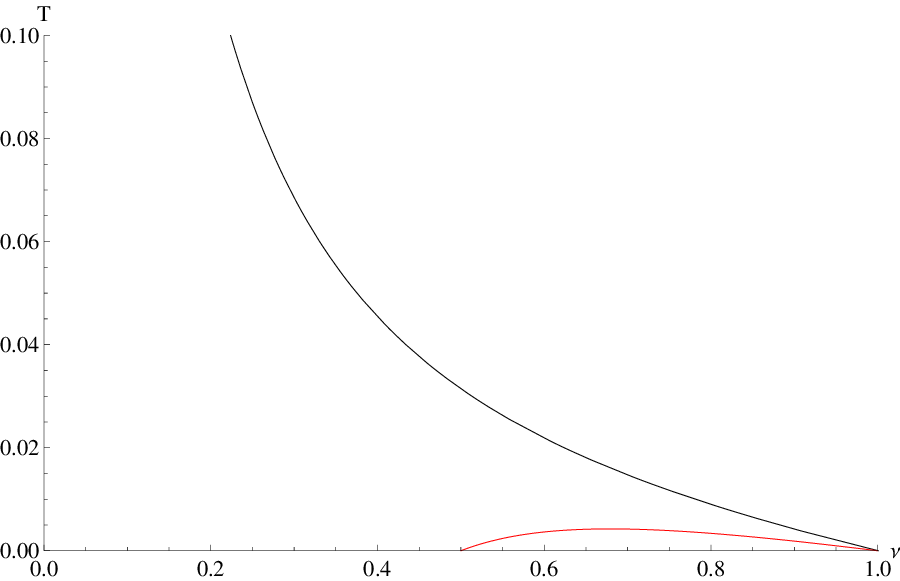}
\caption{{\footnotesize  The standard (black) and modified temperature (red) versus the dimensionless parameter $\nu$ for a spinning black ring.}}
\end{figure}

By making use of the mass $M$, angular velocity $\Omega$ and angular momentum $J$ given by \cite{KERR}
\begin{eqnarray}
M = \frac{3\pi R^2 \nu}{2(1-\nu)(1+\nu^2)}, \quad \Omega = \frac{1}{R}\left[\frac{1-\nu+\nu^2-\nu^3}{2(1+\nu)}\right]^{\frac{1}{2}}, \quad J = \frac{\pi \nu R^3}{\sqrt{2}} \left[\frac{1+\nu}{(1-\nu)(1+\nu^2)}\right]^{\frac{3}{2}},
\end{eqnarray}
we deduce the minimal mass corresponding to the zero temperature,
\begin{eqnarray}
M_{min}=   \frac{3\pi \eta R^4 {E_p}^2 }{2(R E_p-\eta)(R^2 {E_p}^2+{\eta}^2)}.
\end{eqnarray}
The modified entropy can be obtained from the first law of black hole thermodynamics,
\begin{eqnarray}
S &=& \int \left(\frac{dM}{T}-\frac{\Omega dJ}{T}\right)          \nonumber \\
  &=& \frac{\sqrt{2} \pi^2 \left[2R^3 \nu^3+ 3\eta {E_p}^{-1} R^2 \nu^2 + 6 \eta^2  {E_p}^{-2} R \nu + 6 \eta^3 {E_p}^{-3} \ln(R \nu - \eta {E_p}^{-1}) \right]}{\nu (1-\nu)(1+\nu^2)^\frac{3}{2}}.
\end{eqnarray}
From Figure 11 it is obvious that the entropy goes to zero when $\nu$ is in the order of the Planck length, and even declines to negative. Besides two reasons mentioned in Section 2, another reason why the entropy of the spinning black ring becomes negative  near $\nu \approx 0.55$ should be the approximation $E \approx 1/(\nu R)$ we made above.
\begin{figure}[!htbp]
\centering
\includegraphics[height=6cm]{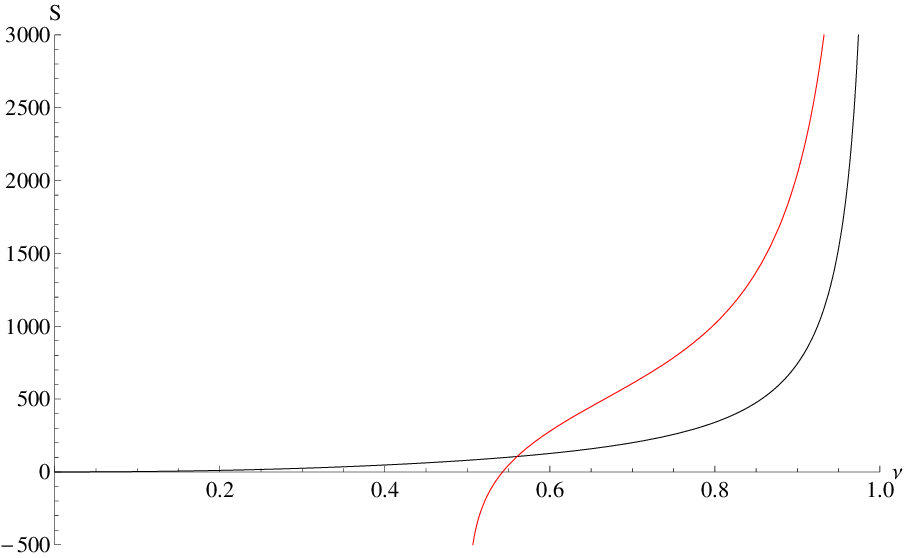}
\caption{\footnotesize The standard (black) and modified entropy (red) versus the dimensionless parameter $\nu$ for a spinning black ring.}
\end{figure}

Similarly, the heat capacity of a spinning black ring at a constant angular momentum is
\begin{eqnarray}
C_J &=& T\left(\frac{\partial S}{\partial T}\right)_J      \nonumber   \\
    &=& \frac{\left(\frac{\partial S}{\partial R}\right)_\nu \left(\frac{\partial J}{\partial \nu}\right)_R - \left(\frac{\partial S}{\partial \nu}\right)_R  \left(\frac{\partial J}{\partial R}\right)_\nu }{\left(\frac{\partial S}{\partial T}\right)_J  \left(\frac{\partial S}{\partial T}\right)_J  -  \left(\frac{\partial S}{\partial T}\right)_J  \left(\frac{\partial S}{\partial T}\right)_J }                                                               \nonumber \\
    &=&   \frac{3 \sqrt{2} \pi^2 K(\nu)}{(1 -  \nu) \nu (1 + \nu^2)^\frac{3}{2} \left[R \nu (2 + 3 \nu^2 + \nu^4) + (-1 + \nu) (4 + \nu + 7 \nu^2 + 4 \nu^3) \eta  {E_p}^{-1} \right]},
\end{eqnarray}
where
\begin{eqnarray}
K (\nu) &\equiv& R \nu \left[2 R^3 \nu^3 (-1 + 2\nu) (1 + \nu^2) - R^2 \nu^2 ( -1 + \nu - 2 \nu^2 + \nu^3 + 5 \nu^4) \eta {E_p}^{-1}   \right.           \nonumber \\
& & \left. - 3 R \nu  ( -1 + \nu - 2 \nu^2 + \nu^3 + 5 \nu^4) \eta^2  {E_p}^{-2}+ 6 ( -1 + \nu - 2 \nu^2 + \nu^3 + 5 \nu^4) \eta^3 {E_p}^{-3} \right]                 \nonumber \\
& & - 6 ( -1 + \nu - 2 \nu^2 + \nu^3 + 5 \nu^4) (R \nu - \eta) \eta^3  {E_p}^{-3}\ln \left(R \nu - \eta  {E_p}^{-1}\right).
\end{eqnarray}
$C_J$ tends to the standard result when $\eta \rightarrow 0 $. Figure 12 shows that the heat capacity of a spinning black ring diverges at the point where the latter reaches its maximum temperature.
\begin{figure}[!htbp]
\centering
\includegraphics[height=6cm]{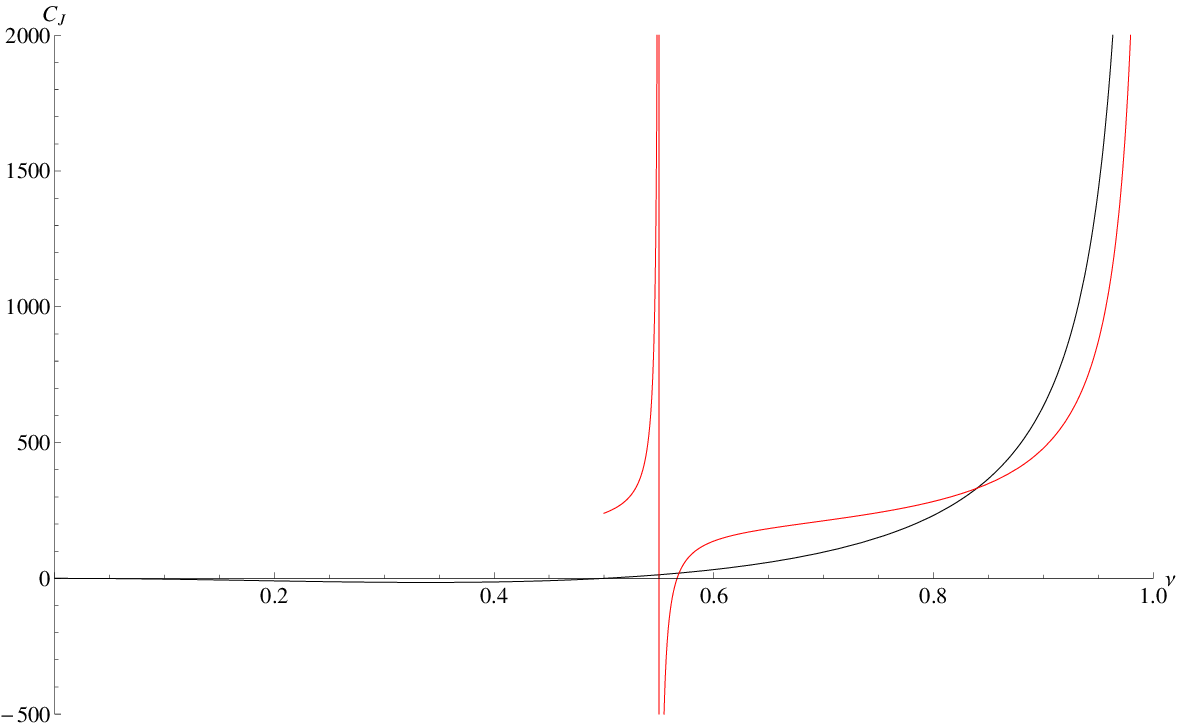}
\caption{{\footnotesize The standard (black) and modified heat capacity (red) versus the dimensionless parameter $\nu$ for a spinning black ring.}}
\end{figure}

%%%%%%%%%%%%%%%%%%%%%%%%%%%%%%%%%%%%%%%%%%%%%%%%%%%%%%%%%%
%%%%%%%%%%%%%%%%%%%%%%%%%%%%%%%%%%%%%%%%%%%%%%%%%%%%%%%%%%
%%%%%%%%%%%%%%%%%%%%%%%%%%%%%%%%%%%%%%%%%%%%%%%%%%%%%%%%%%

\section{Conclusion}

In this paper, we have constructed a new type of gravity's rainbow and investigate its impact on different black holes (black ring). It is found that the new gravity's rainbow function leads to a similar remnant as the appointed rainbow function. We have also derived some important thermodynamic quantities such as the corrected temperature, entropy and heat capacity associated with the Schwarzschild, Kerr, AdS black holes as well as the spinning black ring. The features of these quantities are summarized in the figures above. 

It is evident that at the ending stage of the evaporation process, the black holes (or black ring) do not vanish. On the contrary, such a minimal mass, called black hole remnant, exists and therefore the catastrophic ending of Hawking radiation can be avoided. This indicates that our method is able to reach the equally satisfactory results obtained by the GUP (Generalized uncertainty principle) method. In the framework of GUP, a minimal scale generated from the modification of the Heisenberg relation leads to a black hole remnant as well as corrections to the quantities of black hole thermodynamics. 
Finally, it is worth noting that the formula of the entropy of every black object researched in this paper contains  a  logarithmic term, which leads to a negative entropy near its horizon radius. This problem needs a further discussion in our future work.

\section*{Acknowledgments}
Supported by Research Program of Qilu Institute of Technology (No.: QIT23NN036).

%%%%%%%%%%%%%%%%%%%%%%%%%%%%%%%%%%%%%%%%%%%%%%%%%%%%%%
%%%%%%%%%%%%%%%%%%%%%%%%%%%%%%%%%%%%%%%%%%%%%%%%%%%%%%
%%%%%%%%%%%%%%%%%%%%%%%%%%%%%%%%%%%%%%%%%%%%%%%%%%%%%%

\end{document}